\documentclass{iau} 
\usepackage{epsfig}
\usepackage{graphicx}

\title[Circumstellar structures around HMXBs] 
{Circumstellar structures around high-mass X-ray binaries}

\author[Vasilii V. Gvaramadze]   
{Vasilii V. Gvaramadze$^{1,2}$
}

\affiliation{ $^1$Sternberg Astronomical Institute, Lomonosov Moscow State University, 
Universitetskij Pr. 13, Moscow 119992, Russia \\ email: {\tt vgvaram@mx.iki.rssi.ru}
\\[\affilskip] $^2$Space Research Institute, Russian Academy of Sciences, Profsoyuznaya 84/32, 117997 Moscow, Russia}

\pubyear{2018}
\volume{346}  
\jname{High-Mass X-ray Binaries:
illuminating the passage from massive binaries
to merging compact objects}
\editors{L.~Oskinova, E.~Bozzo, D.~Gies, \& D.~Holz, eds.}
\begin{document}

\maketitle

\begin{abstract}
Many high-mass X-ray binaries (HMXBs) are runaways. Stellar wind and radiation of donor stars in 
HMXBs along with outflows and jets from accretors interact with the local interstellar medium and
produce curious circumstellar structures. Several such structures are presented and discussed in 
this contribution. 
\end{abstract}

\firstsection 
\section{Introduction}

\begin{figure}[b]
\begin{center}
 \includegraphics[width=5.2in]{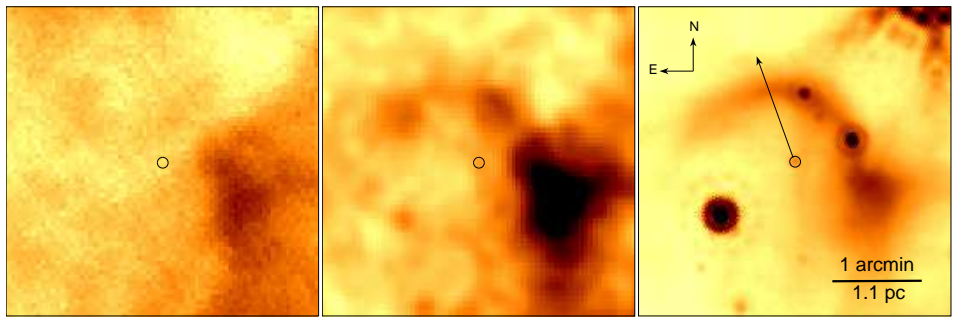} 
 \caption{From left to right: {\it Herschel} 160 and 70\,$\mu$m, and {\it Spitzer} 24\,$\mu$m 
 images of the field containing 4U\,1907+09 (indicated by a circle). The arrow shows the direction 
 of motion of 4U\,1907+09, as follows from the {\it Gaia} proper motion measurement.}
   \label{fig1}
\end{center}
\end{figure}

The high space velocities of HMXBs could be revealed via measurement of proper motions
and/or radial velocities of these systems, or through the detection of bow shocks -- the secondary 
attributes of runaway systems. The first detection of a bow shock produced by a HMXB was 
reported by \cite[Kaper \etal\ (1997)]{Kaper_etal97}, who discovered an H$\alpha$ arc 
around Vela\,X-1. Later, \cite[Huthoff \& Kaper (2002)]{Huthoff02} searched 
for bow shocks around eleven high-velocity HMXBs using {\it IRAS} maps, but did not find new ones. 

Our search for bow shocks around HMXBs from the sample of \cite[Huthoff \& Kaper (2002)]{Huthoff02} 
using data from the {\it Spitzer Space Telescope} led to the discovery of a bow shock associated with
4U\,1907+09 (Fig.\,\ref{fig1}; cf. \cite[Gvaramadze \etal\ 2011]{Gvaramadze_etal11}). An asymmetric
shape of the bow shock could be caused by density inhomogeneities in the local interstellar medium (ISM),
as evidenced by the {\it Herschel} images of the field around 4U\,1907+09 (see Fig.\,\ref{fig1}). The runaway 
nature of this HMXB is supported by proper motion measurements. Particularly, 
the {\it Gaia} DR2 (\cite[Gaia Collaboration 2018]{Gaia Collaboration18}) proper motion and 
distance ($\approx4$ kpc; \cite[Bailer-Jones \etal\ 2018]{Bailer-Jones18}) of 4U\,1907+09 
indicate that this system has a peculiar (transverse) velocity of $\approx200 \, {\rm km} \, 
{\rm s}^{-1}$, which is the highest peculiar velocity measured for HMXBs.

\section{EXO\,1722-363, HD\,34921 \& GX\,304-01}

\begin{figure}[b]
\begin{center}
 \includegraphics[width=4in]{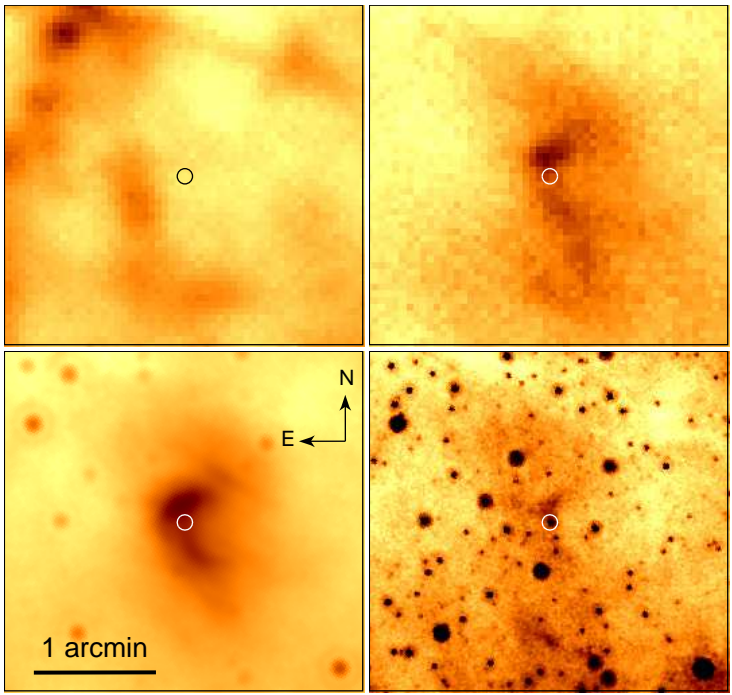}
 \caption{From left to right and from top to bottom: {\it Herschel} 160 and 70\,$\mu$m, and {\it Spitzer} 
 24 and 8\,$\mu$m images of the field containing EXO\,1722-363 (indicated by a circle).}
   \label{fig2}
\end{center}
\end{figure}

We also searched for bow shocks around other HMXBs covered by {\it Spitzer} but, surprisingly, did 
not find any. Instead, we detected curious infrared nebulae around several HMXBs, two of which 
are presented below (both were independently discovered by \cite[Pri\v{s}egen 2018]{Pri\v{s}egen18}).
Fig.\,\ref{fig2} shows a tau-shaped nebula associated with EXO\,1722-363. The shape of the nebula and 
position of EXO\,1722-363 within it exclude the bow shock interpretation for this nebula. Although 
one cannot exclude the possibility that the nebula is produced by collimated outflows (jets) from this 
HMXB (cf. \cite[Gallo \etal\ 2005]{Gallo05}; \cite[Heinz \etal\ 2008]{Heinz08}), the more plausible 
explanation is that we deal with a local ISM heated by radiation from the B0--1\,Ia (\cite[Mason 
\etal\ 2009]{Mason09}) companion star in EXO\,1722-363.   

\begin{figure}[b]
\begin{center}
 \includegraphics[width=4.8in]{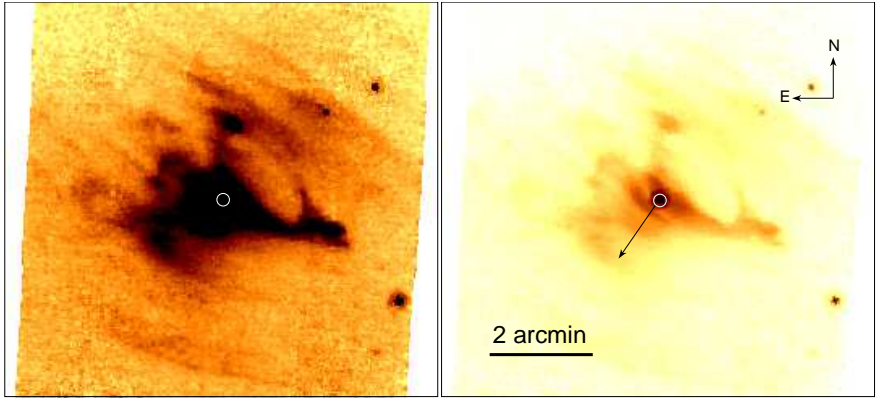}
 \caption{{\it Spitzer} 24\,$\mu$m image of the barrel-like nebula around HD\,34921 (indicated by a 
 circle) in two intensity scales. The arrow shows the direction of motion of HD\,34921.}
   \label{fig3}
\end{center}
\end{figure}

Fig.\,\ref{fig3} shows the {\it Spitzer} 24\,$\mu$m image of a barrel-like nebula around HD\,34921 
in two intensity scales to highlight some details of its filamentary structure. From the {\it Gaia} data 
and the heliocentric radial velocity of HD\,34921 of $-20.5 \, {\rm km} \, {\rm s}^{-1}$ (\cite[Gontcharov 
2008]{Gontcharov06}), we derive the peculiar velocity of this star of $\approx30 \, {\rm km} \, 
{\rm s}^{-1}$. Although this velocity is typical of runaway stars, the complex shape of the nebula excludes 
its interpretation as a pure bow shock.

\begin{figure}[b]
\begin{center}
 \includegraphics[width=4in]{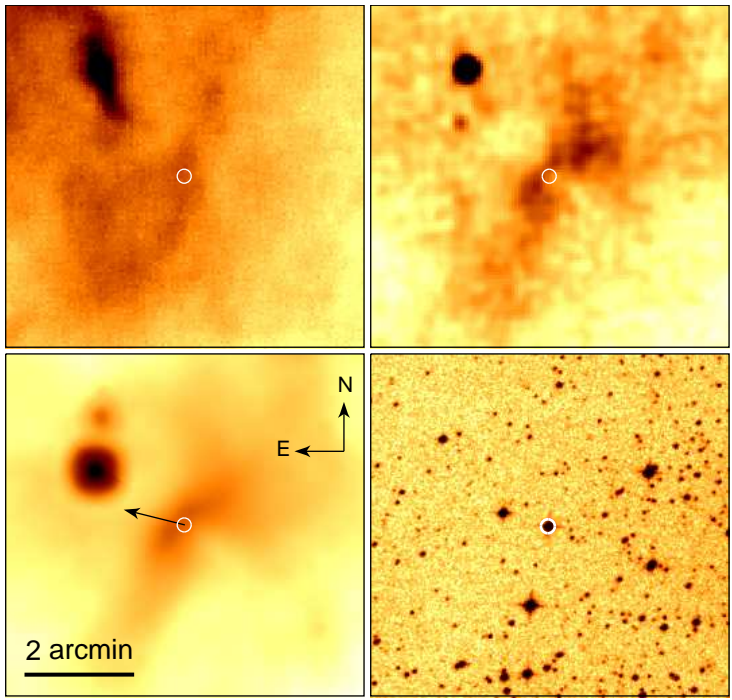} 
 \caption{From left to right and from top to bottom: {\it Herschel} 160 and  70\,$\mu$m, {\it WISE} 
 22\,$\mu$m and DSS-II red-band images of the field containing GX\,304-01 (indicated by a circle). 
 The arrow shows the direction of motion of GX\,304-01.}
   \label{fig4}
\end{center}
\end{figure}

With the advent of the {\it Wide-field Infrared Survey Explorer} ({\it WISE}), it became possible 
to search for bow shocks around all ($\sim100$; e.g. \cite[Liu \etal\ 2006]{Liu_etal06}) known 
HMXBs, and what is absolutely amazing is that we did not find new bow shocks at all! The only 
interesting discovery is a bow-like structure attached to GX\,304-1 (independently detected 
by \cite[Pri\v{s}egen 2018]{Prisegen18}). The geometry of this structure (see Fig.\,\ref{fig4}) and 
the weak wind of the B2\,Vne (\cite[Parkes \etal\ 1980]{Parkes80}) donor star in GX\,304-1 suggest 
that here we deal with a radiation-pressure-driven bow wave (cf. \cite[van Buren \& McCray 
1988]{van Buren88}; \cite[Ochsendorf \etal\ 2014]{Ochsendorf14}), although other explanations 
(involving jets or illumination of the local ISM) cannot be excluded as well.

\section{Vela\,X-1}

\begin{figure}[b]
\begin{center}
 \includegraphics[width=3in]{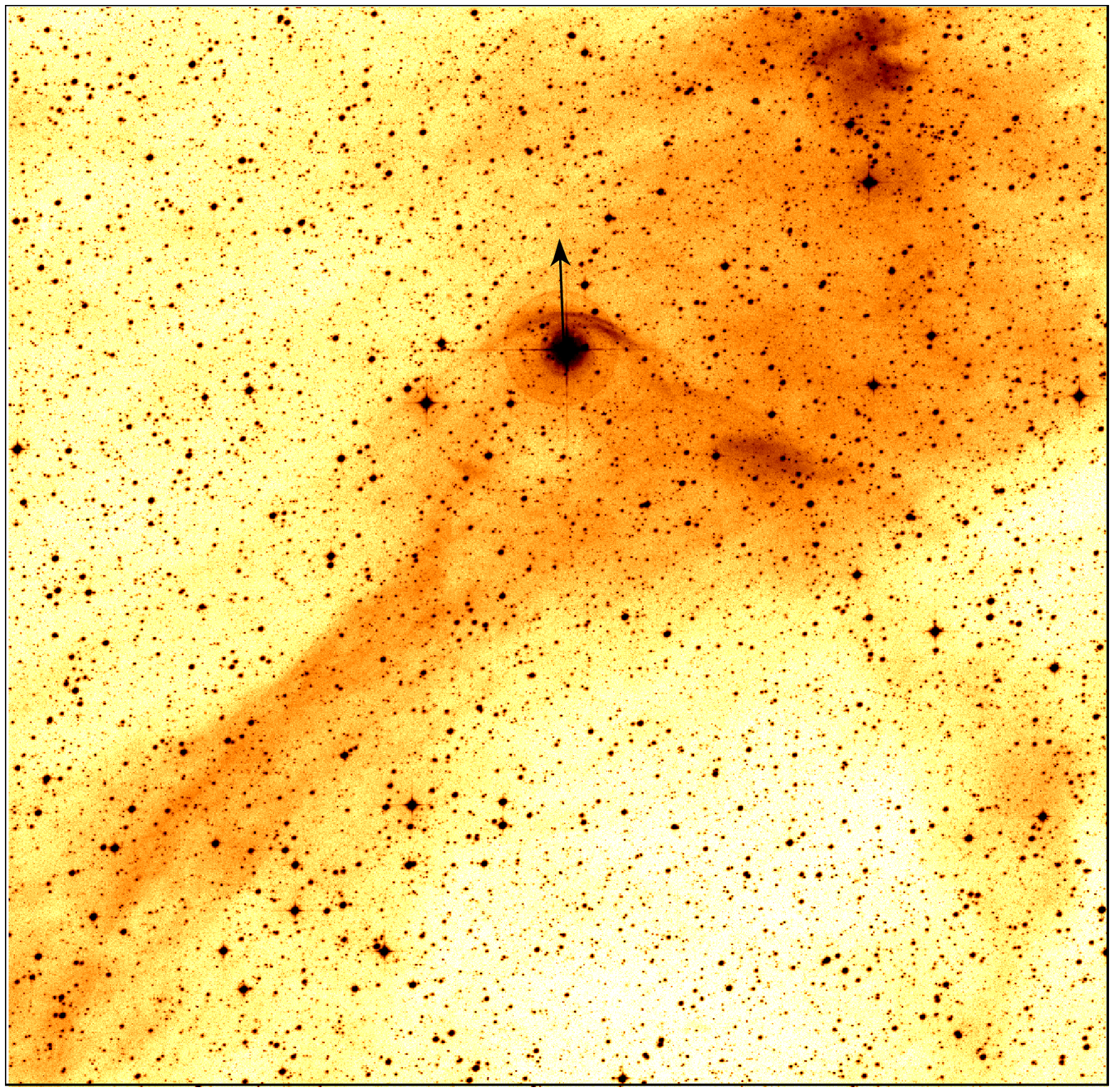}
 \caption{SHS H$\alpha$ image of a 30 arcmin$\times$30 arcmin field containing Vela\,X-1 and 
 filamentary structures behind it. The arrow shows the direction of motion of Vela\,X-1.}
   \label{fig5}
\end{center}
\end{figure}

We also discovered a filamentary structure stretched behind the high-velocity ($\approx50 \, 
{\rm km} \, {\rm s}^{-1}$) HMXB Vela\,X-1. Fig.\,\ref{fig5} shows the SuperCOSMOS H-alpha Survey 
(SHS; \cite[Parker \etal\ 2005]{Parker_etal05}) H$\alpha$ image of this structure along with 
the already known bow shock ahead of Vela\,X-1. The geometry of the filaments suggests 
that Vela\,X-1 has met a wedge-like layer of enhanced density on its way and that the shocked 
material of this layer outlines a wake downstream of Vela\,X-1. 

\begin{figure}[b]
\begin{center}
 \includegraphics[width=5in]{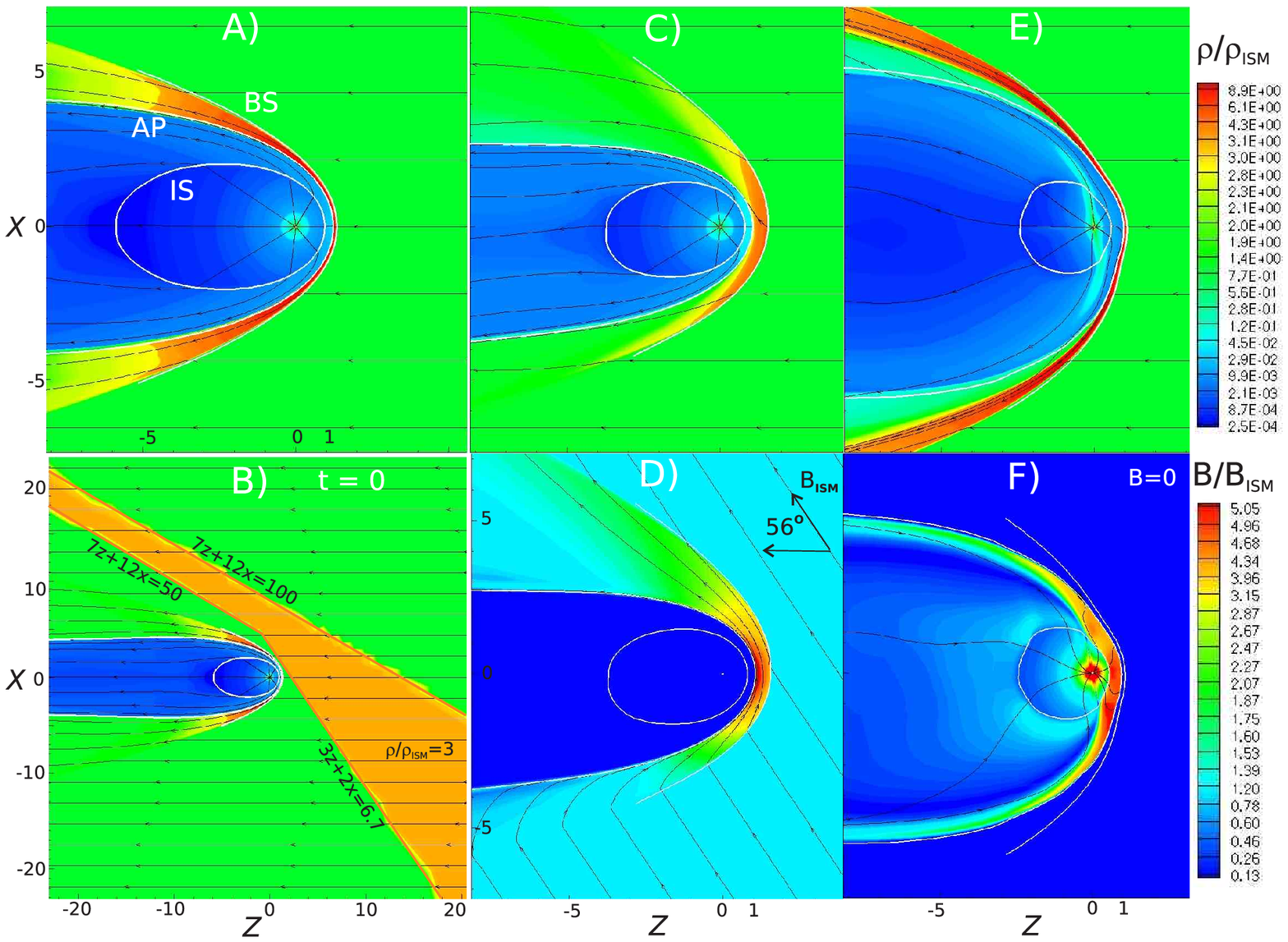} 
 \caption{2D distributions of the plasma density and streamlines (panels A, C, and E) 
and the magnetic field with the field lines (panels D and F) in the steady-state models. 
Panel A corresponds to model 1, panels C and D to model 2, and panels E and F to model 3. 
Panel B shows the initial condition in the ISM for the non-stationary models (shown in
Fig.\,\ref{fig7}). The inner shock (IS), the astropause (AP) and the bow shock (BS) are 
plotted with white lines. Adopted from \cite[Gvaramadze \etal\ (2018a)]{Gvaramadze_etal18a}.}
   \label{fig6}
\end{center}
\end{figure}

\begin{figure}[b]
\begin{center}
 \includegraphics[width=3.5in]{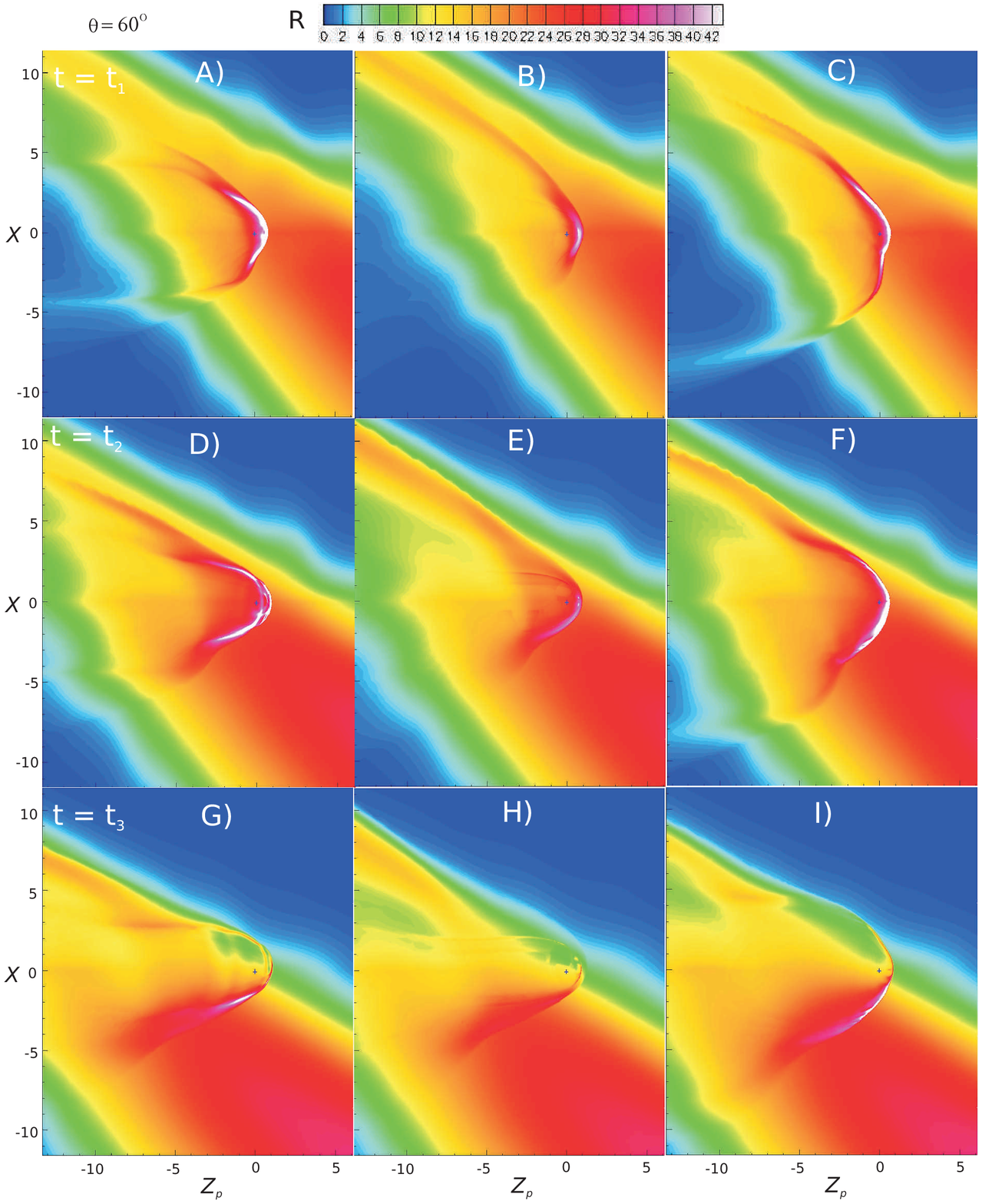}
 \caption{Projection of synthetic H$\alpha$ intensity maps with a line of sight at an angle 
 of $\theta=60^\circ$ to the symmetry axes of the non-stationary models 1, 2, and 3 (left to 
 right) at three times (top to bottom). Adopted from \cite[Gvaramadze \etal\ 
 (2018a)]{Gvaramadze_etal18a}.}
   \label{fig7}
\end{center}
\end{figure}

To substantiate this suggestion, we carried out 3D MHD simulations of interaction between
Vela\,X-1 and the layer for three limiting cases (see Fig.\,\ref{fig6}): the stellar wind and the 
ISM were treated as pure hydrodynamic flows (model\,1); a homogeneous magnetic field
was added to the ISM, while the stellar wind was assumed to be unmagnetized (model\,2); the 
stellar wind was assumed to possess a helical magnetic field (described by the Parker solution; 
\cite[Parker 1958]{Parker58}), while there was no magnetic field in the ISM (model\,3). We found 
that although the first two models can provide a rough agreement with the observations 
(cf. Fig.\,\ref{fig7} with Fig.\,\ref{fig5}), only the third one allowed us to reproduce not only 
the wake behind Vela X-1, but also the opening angle of the bow shock and the apparent detachment 
of its eastern wing from the wake (for more details see \cite[Gvaramadze \etal\ 
2018a]{Gvaramadze_etal18a}).

\section{IGR\,J16327$-$4940}

\cite[Masetti \etal\ (2010)]{Masetti_etal10} detected an OB star within the error circle of 
the {\it INTEGRAL} transient source of hard X-ray emission IGR\,J16327$-$4940 (\cite[Bird 
\etal\ 2010]{Bird10}) and classified this source as a HMXB because of its `overall early-type 
star spectral appearance, which is typical of this class of objects'. 

\begin{figure}[b]
\begin{center}
\includegraphics[width=5.2in,angle=0]{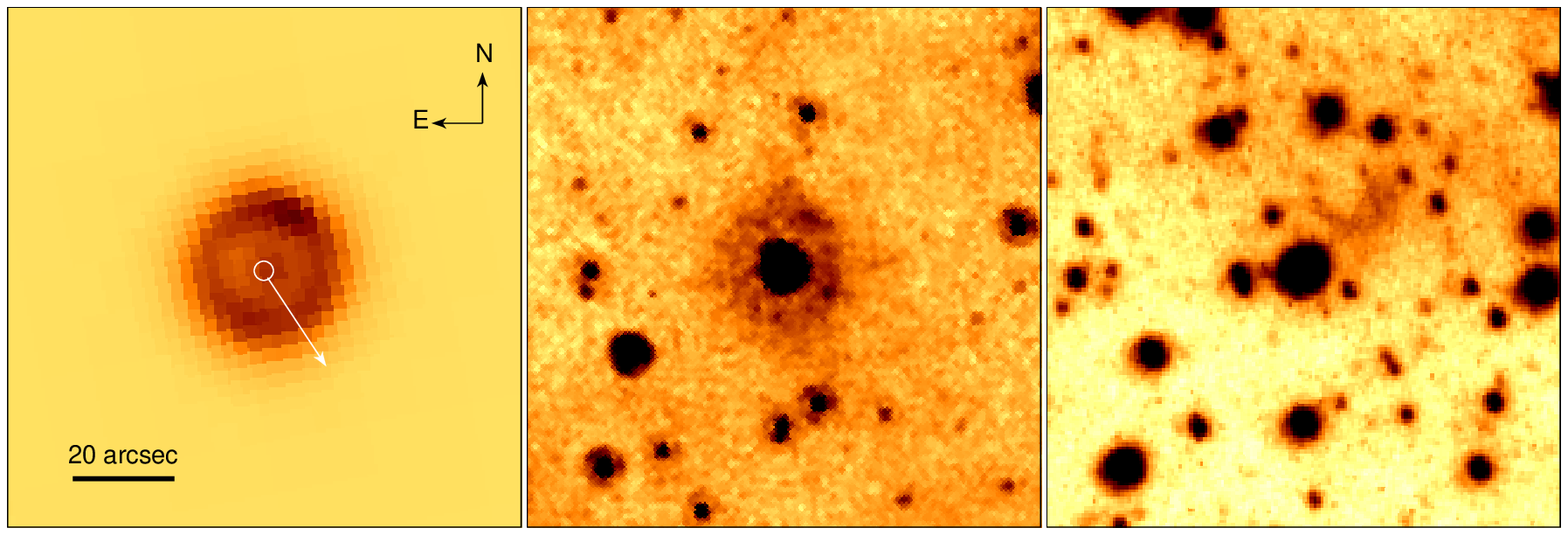}
 \caption{From left to right: {\it Spitzer} 24 and 8\,$\mu$m, and SHS H$\alpha$ images of the 
 field containing IGR\,J16327$-$4940 (indicated by a circle) and its circumstellar nebula. The arrow 
 shows the direction of motion of IGR\,J16327$-$4940.
 }
   \label{fig8}
\end{center}
\end{figure}

Using {\it Spitzer} data, we found that the optical counterpart to IGR\,J16327$-$4940 is 
surrounded by a circular nebula (see Fig.\,\ref{fig8}), named MN44 in \cite[Gvaramadze \etal\ 
(2010)]{Gvaramadze_etal10}. A spectrum of the central star of MN44 taken in 2009 shows hydrogen 
and iron lines in emission, which is typical of luminous blue variables (LBVs) near the visual 
maximum (\cite[Gvaramadze \etal\ 2015]{Gvaramadze_etal15}). New observations carried out in 2015 
revealed significant changes in the spectrum, indicating that the star became hotter. The spectral 
variability was accompanied by $\approx1.6$ mag changes in the brightness, meaning that 
IGR\,J16327$-$4940 is a bona fide LBV (\cite[Gvaramadze \etal\ 2015]{Gvaramadze_etal15}).

{\it Gaia} DR2 data indicate that IGR\,J16327$-$4940 is a high-velocity ($\approx75 \, {\rm km} \, 
{\rm s}^{-1}$) runaway system ejected $\approx4-5$ Myr ago from one of the most massive star clusters 
in the Milky Way --- Westerlund\,1 (\cite[Gvaramadze 2018]{Gvaramadze18}), while a perfectly circular 
shape of the associated nebula implies that it does not feel the effect of ram pressure of the ISM. 
This means that the stellar wind still interacts with a co-moving dense material lost by the star 
during the preceding (e.g. red supergiant) evolutionary stage (cf. \cite[Gvaramadze \etal\ 
2009]{Gvaramadze_etal09}) or because of binary interaction processes (if this star is or was a binary
system; cf. \cite[Gvaramadze \etal\ 2018b]{Gvaramadze_etal18b}). If the HMXB nature of IGR\,J16327$-$4940 
will be confirmed, then this system would represent a first known example of an HMXB with an LBV donor star.

\section{Conclusions}

A possible explanation of the non-detection of bow shocks around HMXBs is that most of these
systems are moving through a low-density, hot medium, so that the emission measure of their bow 
shocks is below the detection limit or the bow shocks do not form at all because the sound speed 
in the local ISM is higher than the stellar peculiar velocity (\cite[Huthoff \& Kaper 
2002]{Huthoff02}). This provides a reasonable explanation of why only one-fifth of runaway OB 
stars produce (observable) bow shocks (\cite[van Buren \etal\ 1995]{van Buren_etal95}). The 
detection rate of bow shocks around HMXBs is, however, a factor of ten less than that for OB stars 
(cf. \cite[Pri\v{s}egen 2018]{Pri\v{s}egen18}). This difference could be understood if the HMXBs have 
systematically lower space velocities compared to the ordinary runaway stars (ejected in the
field mostly because of dynamical few-body interactions), which could be connected to the 
formation mechanism of HMXBs (e.g. the supernova explosions in the HMXB progenitors  
should not be too energetic to unbind them). Moreover, the large proportion of HMXBs with (weak-wind) 
Be donor stars (\cite[Coleiro \etal\ 2013]{Coleiro13}) could also contribute to this difference. 

\vskip 5mm

This work was supported by the Russian Science Foundation grant No. 14-12-01096.


\begin{thebibliography}{}

\bibitem[Bailer-Jones \etal\ (2018)]{Bailer-Jones18}{Bailer-Jones, C.A.L., Rybizki J., Fouesneau, M., 
Mantelet, G., \& Andrae, R.} 2018, \textit{AJ}, 156, 58

\bibitem[Bird \etal\ (2010)]{Bird10}{Bird A.J., {\it et al.}} 2010, \textit{ApJS}, 186, 1

\bibitem[Coleiro \etal\ (2013)]{Coleiro13}{Coleiro, A., Chaty, S., Zurita Heras, J.A., Rahoui, F., 
\& Tomsick, J.A.} 2013, \textit{A\&A}, 560, A108

\bibitem[Gaia Collaboration (2018)]{Gaia Collaboration18}{Gaia Collaboration Brown, A.G.A., 
Vallenari, A., Prusti, T., de Bruijne, J.H.J., Babusiaux, C., \& Bailer-Jones, C.A.L.}  
2018, \textit{A\&A}, 616, A1

\bibitem[Gallo \etal\ (2005)]{Gallo05}{Gallo, E., Fender, R., Kaiser, C., Russell, D., Morganti, R.,
Oosterloo, T., \& Heinz S.} 2005, \textit{Nature}, 436, 819

\bibitem[Gontcharov (2008)]{Gontcharov06}{Gontcharov, G.A.} 2006, \textit{Astron. Lett.}, 32, 759

\bibitem[Gvaramadze (2018)]{Gvaramadze18}{Gvaramadze, V. V.} 2018, {\it RNAAS}, in press; preprint arXiv:1811.07899

\bibitem[Gvaramadze \etal\ (2009)]{Gvaramadze_etal09}{Gvaramadze, V. V., {\it et al.}} 
2009, \textit{MNRAS}, 400, 524

\bibitem[Gvaramadze \etal\ (2010)]{Gvaramadze_etal10}{Gvaramadze, V. V., Kniazev, A. Y.,
\& Fabrika, S.} 2010, \textit{MNRAS}, 405, 1047

\bibitem[Gvaramadze \etal\ (2011)]{Gvaramadze_etal11}{Gvaramadze, V. V., R\"oser, S., 
Scholz, R.-D., \& Schilbach, E.} 2011, \textit{A\&A}, 529, A14

\bibitem[Gvaramadze \etal\ (2015)]{Gvaramadze_etal15}{Gvaramadze, V. V., Kniazev, A. Y., 
\& Berdnikov, L. N.} 2015, \textit{MNRAS}, 454, 3710

\bibitem[Gvaramadze \etal\ (2018a)]{Gvaramadze_etal18a}{Gvaramadze, V. V., Alexashov, D. B.,
Katushkina, O. A., \& Kniazev, A. Y.} 2018a, \textit{MNRAS}, 474, 4421

\bibitem[Gvaramadze \etal\ (2018b)]{Gvaramadze_etal18b}{Gvaramadze, V. V., Maryeva, O. V., Kniazev, A. Y.,
Alexashov, D. B., Castro, N., Langer, N., \& Katkov, I. Y.} 2018b, \textit{MNRAS}, in press; arXiv:1810.12916

\bibitem[Heinz \etal\ (2008)]{Heinz08}{Heinz, S., Grimm, H.J., Sunyaev, R.A., \&
Fender, R.P.} 2008, \textit{ApJ}, 686, 1145

\bibitem[Huthoff \& Kaper (2002)]{Huthoff02}{Huthoff, F., \& Kaper, L.} 2002, \textit{A\&A},
383, 999

\bibitem[Kaper \etal\ (1997)]{Kaper_etal97}{Kaper, L., van Loon, J. Th., Augusteijn, T.,
Goudfrooij, P., Patat, F., Waters, L. B. F. M., \& Zijlstra, A. A.} 1997, \textit{ApJ}  
(Letters), 5475, L37

\bibitem[Liu \etal\ (2006)]{Liu_etal06}{Liu, Q.Z., van Paradijs, J., \& van den Heuvel, E.P.J.} 
2006, \textit{A\&A}, 455, 1165

\bibitem[Masetti \etal\ (2010)]{Masetti_etal10}{Masetti, N., {\it et al.}} 2010, 
\textit{A\&A}, 519, A96 

\bibitem[Mason \etal\ (2009)]{Mason09}{Mason, A.B., Clark, J.S., Norton, A.J., Negueruela, I.,
\& Roche, P.} 2009, \textit{A\&A}, 505, 281

\bibitem[Ochsendorf \etal\ 2014]{Ochsendorf14}{Ochsendorf, B.B., {\it et al.}}
2014, \textit{A\&A}, 563, A65

\bibitem[Parker (1958)]{Parker58}{Parker, E.N.} 1958, \textit{ApJ}, 128, 664

\bibitem[Parker \etal\ (2005)]{Parker_etal05}{Parker, Q., {\it et al.}} 2005, \textit{MNRAS},
362, 689

\bibitem[Parkes \etal\ (1980)]{Parkes80}{Parkes, G.E., Murdin, P.G., \& Mason, K.O.} 1980, 
\textit{MNRAS}, 190, 537

\bibitem[Prisegen (2018)]{Prisegen18}{Pri\v{s}egen, M.} 2018, \textit{A\&A}, in press; arXiv:1811.06781

\bibitem[van Buren \& McCray (1988)]{van Buren88}{van Buren, D., \& McCray, R.} 1988, 
\textit{ApJ}, 329, L93
 
\bibitem[van Buren \etal\ (1995)]{van Buren_etal95}{van Buren, D., Noriega-Crespo, A., 
\& Dgani, R.} 1995, \textit{AJ}, 110, 2914

\end{thebibliography}
\end{document}